\documentclass[reprint,aps,showkeys,pra, nofootinbib, showpacs, floatfix, notitlepage, superscriptaddress, onecolumn]{revtex4-1}
\usepackage[dvipsnames]{xcolor}
\usepackage{subcaption}
\usepackage{hyperref}
\usepackage{braket}
\usepackage{amsmath}
\usepackage{graphicx} 
\definecolor{darkblue}{RGB}{0,0,149}
\hypersetup{
    colorlinks=true,
    linkcolor=red,
    filecolor=darkblue,
    urlcolor=darkblue,
    citecolor=Green,
    linktocpage=true,
}
\usepackage{url}

\usepackage{soul}

\definecolor{coralred}{rgb}{1.0, 0.25, 0.25}

\begin{document}

\title{\ \ \ \ \ \ \ \ \ Quantum effects in gravity beyond the Newton potential   \newline from a delocalised quantum source}
\author{Lin-Qing Chen}%
 \email{linqing.nehc@gmail.com}
\affiliation{Institute for Theoretical Physics, ETH Z{\"u}rich, 8093 Z{\"u}rich, Switzerland}

\author{Flaminia Giacomini}%
 \email{fgiacomini@phys.ethz.ch}
\affiliation{Institute for Theoretical Physics, ETH Z{\"u}rich, 8093 Z{\"u}rich, Switzerland}

\begin{abstract}
    Recent progress in table-top experiments offers the opportunity to show for the first time that gravity is not compatible with a classical description. In all current  experimental proposals, such as the generation of gravitationally induced entanglement between two quantum sources of gravity, gravitational effects can be explained with the Newton potential, namely in a regime that is consistent with the weak-field limit of general relativity and does not probe the field nature of gravity.  Hence, the Newtonian origin of the effects is a limitation to the conclusions on the nature of gravity that can be drawn from these experiments.
    
    Here, we identify two effects that overcome this limitation: they cannot be reproduced using the Newton potential and are independent of graviton emission. First, we show that the interaction between two generic quantum sources of gravity, e.g.\,in  wide Gaussian states, cannot be reproduced with the Newton potential nor with a known classical theory or gravity. Hence, observing the form of this interaction would require either a modification to classical gravity or its quantum description. Second, we show that the quantum commutator between the gravitational field and its canonically conjugate momentum appears as an additional term in the relative phase of a generic quantum source interacting with a test particle. Observing this term in the phase would be a test of the gravitational field as a quantum mediator.

    Identifying stronger quantum aspects of gravity than those reproducible with the Newton potential is crucial to prove the nonclassicality of the gravitational field and to plan a new generation of experiments testing quantum aspects of gravity in a broader sense than what proposed so far.

\end{abstract}

\maketitle

\section{Introduction}

Understanding the fundamental nature of gravity, and in particular whether we have to abandon its classical description, is one of the deepest open questions in fundamental physics. Until recently, this question has mostly been addressed theoretically, because it was not conceivable to devise an experiment that would give a different outcome depending on whether gravity is classical or quantum. However, with the tremendous progress in quantum technologies, it is now possible to plan a new generation of table-top experiments that could measure, for the first time, a physical effect that cannot be explained by any classical theory of gravity.

Starting from a thought experiment proposed by Richard Feynman at the Chapel-Hill conference in 1957~\cite{cecile2011role, zeh2011feynman}, there has been a long debate~\cite{Eppley:1977emg, Page:1981aj, ford1982gravitational, penrose1996gravity, lindner2005testing, kafri2013noise, kafri2014classical, bose2017spin, marletto2017gravitationally, altamirano2018gravity, anastopoulos2015probing, anastopoulos2020quantum, carlesso2017cavendish, bahrami2015gravity, hall2018two, belenchia2018quantum, Qvarfort:2018uag, belenchia2019information, christodoulou2019possibility, howl2020testing, marshman2020locality,Liu:2020zyo, galley2020no, krisnanda2020observable, pal2021experimental, Carney:2021yfw, Carney:2021vvt, Tilly:2021qef, danielson2021gravitationally, kent2021testing, Christodoulou:2022vte, Huggett:2022uui, Chen:2022wro, Christodoulou:2022knr, Danielson:2022tdw, Martin-Martinez:2022uio, Yant:2023smr, Howl:2023xtf, Galley:2023byb, Lami:2023gmz, Schut:2023hsy, Schut:2023eux, Bengyat:2023hxs, Fujita:2023pia} on the possibility to entangle two massive quantum systems via their gravitational interaction. The core of the argument, already discussed at Chapel Hill, is the following: if gravity is a quantum interaction, it can entangle the quantum systems; if it is a classical interaction, it cannot.

The opposite implication does not logically follow. More explicitly, if we observe the generation of \emph{Gravitationally Induced Entanglement} (GIE) in an experiment, we cannot immediately conclude that gravity is quantum, for several reasons. For instance, if entanglement is generated via the Newton potential, we need to further assume that gravity acts as a mediator\cite{bose2017spin, marletto2017gravitationally, galley2020no}. However, it is in general subtle to prove in an experiment which gravitational degrees of freedom are responsible for the entanglement. Second, from a logical point of view we cannot exclude that the experimental results can be justified by some other description that does not require the quantisation of the gravitational field. This problem is analogous to the debate, that started with the discovery of the photoelectric effect, concerning which observation would be a convincing proof that electromagnetism is quantum. Notably, this debate was only closed after over half a century of discussion, thanks to an experiment by J. Clauser~\cite{Clauser:1974gd} in 1974. To date, there is no agreement on what would constitute an analogously convincing (and realistic) observation for gravity.

Recently, many different arguments have been put forward to specify in which sense observing GIE would imply that gravity is not classical. For instance, if the LOCC theorem (namely, the impossibility to generate entanglement via local operations and classical communication) is applied to GIE then, assuming that gravity mediates the interaction, one can prove that gravity is not classical~\cite{bose2017spin, marletto2017gravitationally, Lami:2023gmz}. Other authors have adopted theory-independent methods~\cite{Marletto:2017pjr, Marletto:2020cdx, galley2020no, Galley:2023byb}, e.g.\,to formulate no-go theorems that, based solely on the experimental outcomes and on some general principles, can exclude a classical description of gravity. In a theory-specific approach, it is possible to use arguments resting on the notion of locality~\cite{Christodoulou:2022knr}, on the reference frame dependent nature of the gauge~\cite{danielson2021gravitationally}, or on the explicit characterisation of the field degrees of freedom~\cite{Chen:2022wro}. Such theory-specific arguments do not prove the quantum nature of gravity, but provide a theoretical framework in which the results can be interpreted.

Despite all the progress achieved in the past few years, there is still a lack of consensus on whether observing GIE is a convincing enough observation that gravity is not classical. Such a lack of consensus derives from the fact that in all proposed experiments~\cite{bahrami2015gravity, anastopoulos2015probing, bose2017spin, marletto2017gravitationally, carlesso2017cavendish, Qvarfort:2018uag, christodoulou2019possibility, howl2020testing, marshman2020locality, krisnanda2020observable, pal2021experimental, Carney:2021yfw, kent2021testing,  Christodoulou:2022vte, Huggett:2022uui, Christodoulou:2022knr, Chen:2022wro, Martin-Martinez:2022uio, Howl:2023xtf, Lami:2023gmz, Bengyat:2023hxs, Fujita:2023pia} all gravitational effects can be explained using solely the Newton potential. The Newton potential is the classical solution of Einstein's equations in the non-relativistic and weak-field regime and it can be used to explain GIE without invoking any quantum feature of gravity as a field~\cite{hall2018two, Ma:2021rve, Fragkos:2022tbm, Martin-Martinez:2022uio, Marchese:2024zfu}. So far, there is no experimental proposal that proves that gravity is quantum without involving additional assumptions on the field nature of gravity, i.e.\,the fact that gravity cannot be reduced to the Newton potential alone.

Regardless of the personal take on this discussion, measuring GIE via the Newton potential would be a seminal result, because it would be the first measurement that cannot be explained using the classical theory of general relativity (as is the case for all experiments performed so far). However, it is now crucial to search for stronger evidence of quantum aspects of gravity that could be tested in future experiments. Identifying richer effects in this regime would have a crucial impact on the scope and relevance of future tests of the quantum nature of gravity: if they exist, this will open a new phenomenological window on quantum effects in gravity. 

Here, we provide for the first time two examples of such more general effects, which are of the same order as the Newton potential in the gravitational coupling. Importantly, these effects are also independent of graviton emission. 
 We explicitly derive the two effects using a field-basis formulation of linearised quantum gravity in the Schr{\"o}dinger representation~\cite{Chen:2022wro}. Note that this is a physical regime~\cite{Donoghue:1994dn, donoghue2017epfl} where it is expected that all non-perturbative quantum gravity theories agree in the low energy limit. In the first case, we consider two quantum sources of gravity prepared in a wide delocalised state, and we show that the gravitational interaction of such sources cannot be reproduced with the Newton potential nor with a known classical theory or gravity. Hence, observing this phase would require either an \emph{ad hoc} modification to classical gravity or its quantum description. In the second case, we consider a simple physical scenario involving a source and a moving test particle, and by carefully analysing all terms coming from the full Hamiltonian description of the interaction between matter and gravity, we show that the quantum commutator between the gravitational field and its canonically conjugate momentum appears in the relative phase accumulated during time evolution. This is in contrast with the traditional expectation that the physical effect coming from the gravitational commutators is only relevant at very high energy scale.   We notice that the phase due to the commutator would not appear if gravity were a classical field. Hence, probing this additional term in the phase would be a test of the gravitational field as a quantum mediator. 

The paper is organised as follows. In Sec.\,\ref{sec:SemiclassicalS}, we spell out the assumptions that are usually taken in the analysis of the GIE proposal, that lead to identifying the Newton potential as the origin of entanglement. In Sec.\,\ref{sec:gravitywidesource}, we show how these assumptions can be abandoned, and we describe the gravitational field associated to a general quantum source. In Sec.\,\ref{sec:widesource}, we calculate the phase arising from the interaction between two quantum sources, and show that it cannot be reproduced via the Newton potential nor via other known semiclassical descriptions of gravity. In Sec.\,\ref{sec:commutator}, we show how the commutator between gravity and its canonically conjugated momentum contribute to the relative dynamical phase of the interaction between a quantum source and a test particle.

\section{Semiclassical localised source and its gravitational field}
\label{sec:SemiclassicalS}

All current proposals for table-top tests of gravity are in a regime in which gravitational effects can be purely explained  by the Newton potential. 
In this Section, we identify the assumptions one needs to make on the matter source to obtain the Newton potential in the linearised regime of an effective field theory of gravity. In the next Section, we explicitly derive the quantum state of the gravitational field associated to a generic quantum source. In addition, we show that when we consider general quantum states of matter sourcing gravity, the physical effects are also more general than those usually obtained with the Newton potential. Finally, we show that one can recover the Newton potential by imposing precise restrictive conditions on the quantum state of matter and gravity:
the Newton potential corresponds to the quantum source being in a minimum dispersion wavepacket (e.g.\,a coherent state), whose delocalisation is smaller than the experimental resolution in both position and momentum, and the gravitational field in the lowest possible energy eigenstate.

In the literature, a localised quantum source is most commonly represented using a position eigenstate. However, this state is not realistically realisable in the laboratory, but it provides a good approximation of a more realistic coherent state in the experimental regimes considered~\cite{Chevalier:2020uvv}. Let us denote such a coherent state as $\ket{\alpha_i}_S$, and consider it to be localised around position $x_i$ and momentum $p_i=0$. For any parameter $\alpha_i = \alpha_i^R + i \alpha_i^I$, the mean position and momentum of the coherent state are related to $\alpha$ via $x_i = \langle \hat{x}_S \rangle_\alpha \propto \alpha_i^R$ and momentum $p_i = \langle \hat{p}_S \rangle_\alpha \propto \alpha_i^I$.

For a static source, the only non-zero component of the energy-momentum tensor $T_{\mu\nu}$ is the $T_{00}$ component. In quantum theory, $\hat{T}_{00}$ is in general a function of both the position and momentum operator, namely $\hat{T}_{00} (\hat{x}_S, \hat{p}_S)$\footnote{This is completely analogous to the charge density  $J_0$ in quantum electrodynamics. A full description of the $\hat{T}_{00}$ operator requires quantum field theory. There, $\hat{T}_{00}$ is the Hamiltonian energy density on the field configuration and its conjugate momentum~\cite{Peskin:1995ev}.}. A semiclassical localised source $| \alpha_i  \rangle_{S} $ is such if it satisfies the condition
\begin{equation} \label{eq:semicl}
\hat{T}_{00} | \alpha_i  \rangle_S   \approx  \rho_i (\vec{x} - \vec{x}_i, t)| \alpha_i  \rangle_S, 
\end{equation}
where $\rho_i (\vec{x} - \vec{x}_i, t)$ is the classical energy density in general relativity. For a static pointlike source, $\rho_i (\vec{x} - \vec{x}_i,t) \propto m c^2 \delta(\vec{x} - \vec{x}_i)$, with $m$ being the rest mass of the source. In particular, Eq.\,\eqref{eq:semicl} is satisfied if
\begin{equation} 
    \hat{x}_S |\alpha_i\rangle_S  \approx  x_i|\alpha_i\rangle_S, \qquad \hat{p}_S |\alpha_i\rangle_S  \approx  p_i|\alpha_i\rangle_S.
\end{equation}
Physically, Eq.\,\eqref{eq:semicl} holds when, for any relevant operational procedure, the commutator between the position and momentum of the source is negligible compared to the precision of the measurement device, and so is the width of the Gaussian in position basis.  Therefore, $\hat{T}_{00}$ is mutually diagonalisable in $\hat{x}$ and $\hat{p}$. Notice that, since we are assuming the source to be static, we do not need any condition on the time evolution of the source.

Under these assumptions, we can calculate the quantum state of gravity $|g_\alpha\rangle$ associated to a coherent state $\ket{\alpha}_S$ in the temporal gauge, i.e.\,$h_{0\mu}=0$ (see App.\,\ref{app:SemClassG} for details). In the Newtonian regime, in which there is no emission of radiation,  the gravitational field needs to be in the ground-state of the linearised quantum gravity Hamiltonian. Following the calculations of Ref.\,\cite{Chen:2022wro}, which we also summarise in App.\,\ref{app:SemClassG}, we find that
\begin{equation}
    \ket{\Psi_\alpha}_{S+G} = \ket{\alpha}_S \ket{g_\alpha}_G,
\end{equation}
where
\begin{equation} \label{eq:Psialpha}
\begin{split}
|\Psi_\alpha\rangle_{S+G} &=  \eta \int \mathcal{D}[h_{ij}] \delta[h^T-\mathsf{h}^T_{\rho_{\alpha } } ] \Psi_{vac} [h_{ij}] \ket{\alpha }_S \ket{h_{ij}}_G \\
&=  \eta' \int \mathcal{D}[\pi_{ij}] \exp \left\{ - \frac{i}{2 \hbar} \int \frac{d^3k}{(2 \pi)^3} \pi_T(\vec k)  \mathrm{h}^T_{\rho}(\vec k) \right\} \Psi_{vac}[\pi_{ij}] \ket{\alpha }_S \ket{\pi_{ij}}_G  
\end{split}
\end{equation} 

In the expressions above, $h_{ij}$ and $\pi_{ij}$ are the metric perturbation and its canonically conjugated momentum, $ h_T^{ij}$ and $\pi_T^{ij}$ are their projection on the transverse direction, i.e.\,$ h_T^{ij} (\vec{k}):=P^i_k P^j_l h^{kl} (\vec{k})$ and $ \pi_T^{ij} (\vec{k}):=P^i_k P^j_l \pi^{kl} (\vec{k})$, with $P^i_j = \delta^i_j - \frac{k^i k_j}{|\vec k|^2}$. Finally, $h^T = \delta_{ij} h_T^{ij}$ and $\pi^T = \delta_{ij} \pi_T^{ij}$. In the representation of the canonical momenta $\pi_{ij}$, the effect of the matter source on the ground state of gravity is a shift by a phase dependent on the solution of the classical Poisson equation $\partial_j \partial^j  h^T(x) = -\kappa\rho_i(x)$~\cite{Chen:2022wro} (we here dropped the $t$ argument in the source because it is static). In the $h_{ij}$ representation, the same condition is imposed via a Dirac delta in the expression of the quantum state.

The functional of the gravitational field $\Psi_{vac}$ is the ground state solution of the source-free gravity Hamiltonian
\begin{equation}
    \begin{split}
        \Psi_{vac}[h_{ij}] &\propto \exp\left\{ - \frac{1}{4 \kappa \hbar} \int \frac{d^3k}{(2\pi)^3}\  |\vec{k}|\    h^T_{ij}(\vec{k})  h_T^{ij} (-\vec{k}) \right\}\\
        \Psi_{vac}[\pi_{ij}]&\propto \exp \left\{ - \frac{\kappa} {\hbar}\int \frac{d^3k}{(2 \pi)^3} \frac{1}{|k|} \left(   \pi^T_{ij}(\vec k)   \pi_T^{ij}(-\vec k) - \frac{1}{2} \pi_T(\vec k)  \pi_T(-\vec k)  \right) \right\}.
    \end{split}
\end{equation} 

Eq.\,\eqref{eq:semicl} is crucial to obtain the Newton interaction for the static sources in this regime. However, that assumption is not justified for general quantum sources. Specifically, if we use Eq.\,\eqref{eq:semicl} for different quantum states of gravity associated to different localised states of the source,  we obtain some quantum features that are not realistic. For instance, let us consider two coherent states of the source, $\ket{\alpha_{x}}_S$ and $\ket{\alpha_{x+\epsilon}}_S$, whose central positions are displaced by an arbitrarily small amount $\epsilon$, with $\epsilon$ smaller than the width of the coherent state of the source. Their scalar product is different to zero, i.e.\,$\braket{\alpha_{x}|\alpha_{x+\epsilon}}\neq 0$. However, the scalar product between the respective full quantum states of matter and gravity, $\ket{\Psi_{x}}_{S+G}$ and $\ket{\Phi_{x+\epsilon}}_{S+G}$ is zero, namely $\braket{\Psi_{x} | \Phi_{x+\epsilon}}_{S+G}=0$ (see App.\,\ref{app:SemClassG} for details). This means that the description of Eq.\,\eqref{eq:semicl} cannot be adopted for general quantum states, and can only describe superpositions of quantum states of the source that are perfectly distinguishable (e.g.\,the configurations used in the typical GIE experiment). Hence, this approximation should be abandoned to explore more non-trivial quantum aspects of gravity.

Any quantum superposition of such a semiclassical localised source leads to an associated gravity configuration that is not classical, but only in a very limited sense. In particular, for a quantum state $\ket{\psi}_S = \sum_i c_i \ket{\alpha_i}_S$, we have
\begin{equation} \label{eq:suplocstate}
  \hat{T}_{00} | \psi \rangle_{S} = \sum_i c_i \rho_i (\vec x-\vec x_i, t)| \alpha_i \rangle_{S}
\end{equation}
The resulting quantum state of the gravitational field is then
\begin{equation} \label{eq:semiclGrav}
    \ket{\Psi_\psi}_{S+G} = \sum_i c_i \ket{\alpha_i}_S \ket{h_i}_G,
\end{equation}
but the difference between any two gravity states $\ket{h_i}_G$ and $\ket{h_j}_G$, for $i\neq j$, is only in $\mathsf{h}^T_{\rho_{\alpha_i}}$ in Eq.\,\eqref{eq:Psialpha}, i.e.\,the solution of the classical Poisson equation for a static source. In this case, all gravitational effects can be simply expressed as a quantum superposition of classical gravitational effects.
This is the usual approximation implicitly adopted, for instance, for a quantum source of gravity prepared in a superposition of localised states, as in GIE proposals~\cite{bose2017spin, marletto2017gravitationally}. 

\begin{figure}
    \centering
    \includegraphics[scale=0.3]{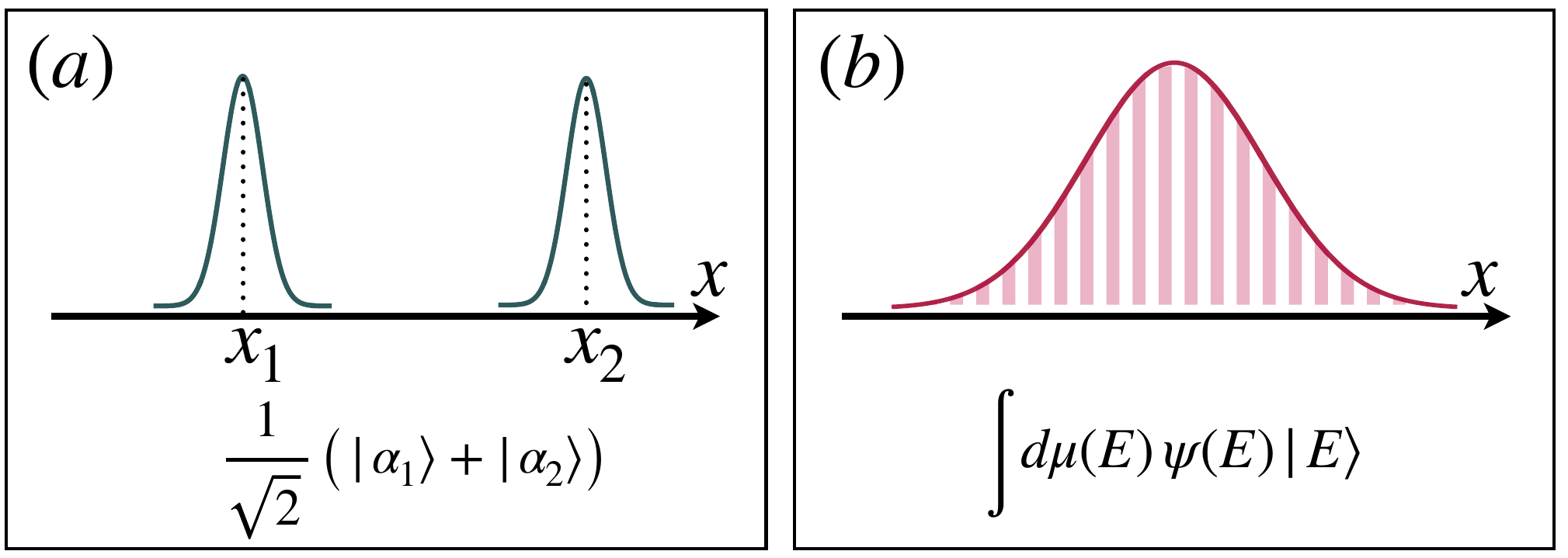}
    \caption{\textbf{(a) Superposition of semiclassical localised source.} The source is prepared in a quantum superposition of two (or more) Gaussian states. Each state is sharply localised around some central value of position $x_i$ and momentum $p_i=0$. The resolution of the measurement devices is such that, for all practical purposes, the delocalisation of the quantum state in both position and momentum is negligible. Under this assumption, each Gaussian quantum state is an eigenvector of the $\hat{T}_{00}$ operator having the classical energy density $\rho(\vec x - \vec x_i, t)$ as its eigenvalue. The quantum state of the gravitational field gives rise to the Newton potential.
\textbf{(b) General delocalised quantum source.} The source is prepared in an arbitrary delocalised quantum state, for simplicity of illustration represented as a Gaussian state in the picture. The quantum state cannot be approximated as a linear combination of semiclassical localised states, but should be solved directly in a convenient basis, i.e. the energy basis $\ket{E}$, which is the basis of eigenvectors of the $\hat{T}_{00}$ operator. The eigenvalues of the $\hat{T}_{00}$ operator do not coincide, in the general case, with the classical energy density, hence the resulting gravitational interaction is functionally different to the Newton potential.}
    \label{fig1}
\end{figure}

\section{General static quantum sources and their gravitational field}
\label{sec:gravitywidesource}

We now show that richer properties of the gravitational field can be obtained by abandoning the condition of Eq.\,\eqref{eq:semicl}. To see this, we consider a general static quantum source of gravity (see Fig.\,\ref{fig1} for an illustration of the difference with the previous Section). The $\hat{T}_{00}$ operator of such a quantum source is, strictly speaking, an operator written as a function of the quantum field of the source.  In particular, it is equal to the Hamiltonian density $\hat{\mathcal{H}}(x)$ of a quantum field, such as a Klein-Gordon field~\cite{Peskin:1995ev}, i.e.\,$\hat{T}_{00}(x) = \hat{\mathcal{H}}(x)$. It is convenient to perform the calculations in a basis that diagonalises $\hat{T}_{00}$, namely $\hat{T}_{00} |E \rangle_S = E(\vec{x}) |E \rangle_S$. Here, the source is described as the low-energy and first-quantised limit of the more fundamental field-theoretic description\footnote{For a  static source, the Schwinger term~\cite{Schwinger:1959xd, Deser:1967zzf} $[\hat{T}_{00}(\vec{x}), \hat{T}_{00}(\vec{x}')] = -i \hbar (\hat{T}^{0k}(\vec{x})+ \hat{T}^{0k}(\vec{x}'))\partial_k \delta (\vec{x}-\vec{x}')$ is negligible. Therefore, $\hat{T}_{00}$ commutes at different spatial locations for all cases we consider.}. To obtain our result, it is only relevant to describe the quantum state of the source in first quantisation and to know that, since the physical interpretation of the $\hat{T}_{00}$ is an energy density, the eigenvalue $E(\vec{x})$ is a local function on spacetime. More specifically, we are interested in the low-energy limit in which the source of gravity is in a static quantum state $\ket{\psi}_S$. This state is more general compared to the coherent states discussed in the previous Section, hence $E(\vec{x})$ has a different functional form to the classical energy density $\rho_i (\vec x-\vec x_i, t)$. A general quantum state of the source can be expanded as 
\begin{equation}
 |\psi \rangle_S = \int d \mu(E) \psi (E)  |E \rangle_S.
\end{equation}
The quantum state of the gravitational field associated to such a source can be obtained with an analogous procedure to the one used in Ref.\,\cite{Chen:2022wro}. We only report the main steps and refer the reader to App.\,\ref{app:DetQuantumSource} for details. The free Hamiltonian of the gravitational field is
\begin{equation}\label{eq:Hgr}
\hat{H}_G = \kappa  \int \frac{d^3k}{(2 \pi)^3}  \  \left(\pi_{kl} \pi^{kl} - \pi^2/2 \right) + \frac{1}{4\kappa}   \int \frac{d^3k}{(2 \pi)^3} \left( k^2  h^T_{ij}(\vec{k})  h_T^{ij}(-\vec{k})  - k^2 h_T(\vec{k}) h_T(-\vec{k})\right).
\end{equation}
in which $ h_T^{ij}$ is the transverse part of the metric perturbation $ h_T^{ij} (\vec{k}):=P^i_k P^j_l h^{kl} (\vec{k}) $ and $P^i_j = \delta^i_j - \frac{k^i k_j}{|\vec k|^2}$.
In addition, the physical state has to satisfy the scalar and vector constraints
\begin{equation} \label{eq:constwide}
    \hat{\mathcal{C}} : = - \partial_i \partial^i \hat{h}^T - \kappa \hat{T}_{00} =0, \qquad \hat{\mathcal{C}}^i : =  \partial_j \hat{\pi}^{ij} =0.
\end{equation}
We notice that the static source only influences the physical state through the scalar constraint $\hat{\mathcal{C}}$. Solving the constraints in the basis $\ket{E}_S$ for the source and in the basis $\ket{\pi_{ij}}$ of the gravitational field, we obtain
\begin{equation}\label{scalarcons}
\partial_i \partial^i \hat{h}^T | \Psi_\psi \rangle_{S+G} = 
 - i \hbar \partial_i \partial^i P_{kl} \frac{\delta}{\delta \pi_{kl}}  | \Psi_\psi \rangle_{S+G} = - \kappa \int d \mu(E) \psi (E) E (t, \vec{x}) |E \rangle_S  | g_E \rangle_G.
\end{equation}
The general solution of Eq.\,\eqref{scalarcons} is  a joint state of the gravitational field and the source, namely
\begin{equation}\label{statepi}
    \begin{split}
        \ket{\Psi_{\psi}}_{S+G} &= \eta \int d\mu(E)\mathcal{D}[h_{ij}] \psi (E) \delta[h^T-\mathsf{h}^T_{E} ]  \Psi_{vac}[h_{ij}] \ket{E}_S \ket{h_{ij}}_G\\
        &= \eta' \int d\mu(E) \mathcal{D}[\pi_{ij}] \psi (E)\exp \left( - \frac{i}{2 \hbar} \int d^3x \pi_T(\vec x) \mathrm{h}_{E}^T(\vec x)\right) \Psi_{vac}[\pi_{ij}] \ket{E}_S\ket{\pi_{ij}}_G,
    \end{split} 
\end{equation}
where $\eta,\,\eta'$ are normalisation constants,  $\mathrm{h}^T_{E}(\vec{x})$ is the solution of the classical  Poisson equation with the source being in the eigenstate $\ket{E}_S$ with eigenvalue $E (\vec x)$. In particular, we have
\begin{equation}\label{deth}
    \mathrm{h}^T_{E}(\vec{x}) = \frac{\kappa}{4 \pi} \int d^3 y \frac{E(\vec y)}{|\vec x - \vec y|}.
\end{equation}

The integration measure of the metric field can be decomposed into a longitudinal part $h_{ij}^L$, a transverse traceless part $\tilde{h}_{ij}^T$, and a transverse trace part $h_T$, namely $\mathcal{D}[h_{ij}]  = \mathcal{D}[h_{ij}^L] \mathcal{D}[\tilde{h}_{ij}^T] \mathcal{D}[h_T]$. 

 Differently to the case of the localised semiclassical source of Sec.\,\ref{sec:SemiclassicalS}, two overlapping quantum states of the source $|\psi \rangle$ and $|\varphi\rangle$ such that $\braket{\psi|\phi}\neq 0$ give rise to quantum states of gravity that are not perfectly distinguishable. This can be easily seen by computing directly the scalar product of the quantum states of gravity and matter (see App.~\ref{app:DetQuantumSource} for details)
\begin{equation}
\langle \Psi_{\psi}|\Psi_{\phi} \rangle_{S+G} = \int d \mu(E) \psi^* (E) \phi (E)  = \langle \psi| \phi \rangle_{S} \neq 0,
\end{equation}
as we would intuitively expect. Interestingly, the scalar product only depends on the states of matter, and does not depend on gravity.

\section{Physical effects beyond the Newtonian phase for a wide quantum source}
\label{sec:widesource}

We now consider the scenario in which two static delocalised quantum sources $A$ and $B$  interact gravitationally.  We derive the interaction between the two masses and show that the relative phase in this scenario cannot be reproduced with the Newton potential. Let us define as $\hat{h}_{\mu\nu}(x)$ the linearised quantum gravitational field sourced by the mass, and as $\hat{T}_{\mu\nu}^{A}$ and $\hat{T}_{\mu\nu}^{B}$ the stress-energy tensors of source $A$ and $B$ respectively. The Hamiltonian is the sum of a free and an interaction term, namely
\begin{equation} \label{eq:fullH}
   \hat{H} = \hat{H}_{A} + \hat{H}_{B}+ \hat{H}^0_G + \hat{H}_I^{(tot)},
\end{equation}
where $\hat{H}_{A}, \hat{H}_{B}$ are the source Hamiltonians and $\hat{H}^0_G$ is the source-free gravitational field Hamiltonian. For the purpose of this section, the explicit expressions of the source Hamiltonians are not needed. The interaction Hamiltonian between $A$, $B$ and gravity is, in its most general form
\begin{equation}
    \hat{H}_I^{(tot)} = - \frac{1}{2}\int d^3x\, \hat{h}_{\mu\nu}(\vec x) (\hat{T}^{\mu\nu}_{A} (\vec x)+ \hat{T}^{\mu\nu}_{B} (\vec x)).
\end{equation}
For  static quantum sources,  $T_{0 0}$ is the only non-zero component of their stress-energy tensors\footnote{Note that, when one transforms  into the frame compatible with the temporal gauge, the stress-energy tensor of the static mass can acquire nontrivial spatial components. However, they are higher order in $G_N$, and therefore negligible in this scenario.}.
Since $h_{0\mu} =0$ in the temporal gauge, this means that the interaction Hamiltonian $   H_I^{(tot)}$ is zero at the leading order in $G_N$ expansion. Crucially, the interaction between the mass source and the gravitational field is fully encoded in the Gauss constraint $\hat{\mathcal{C}} : = - \partial_i \partial^i \hat{h}^T - \kappa \hat{T}^A_{00} - \kappa \hat{T}^B_{00} =0$. We consider the equilibrium state, during which the two sources are relatively static to each other for a period of time t and all the radiation has dissipated away. Expanding both quantum sources in the eigenbasis of $\hat{T}_{00}$, the full quantum state of the gravitational field and the sources is
\begin{equation}
\ket{\Psi}_{ABG} 
        = \eta \int d\mu(E_A) d\mu(E_B)\mathcal{D}[\pi_{ij}] \psi (E_A) \phi (E_B)\exp \left( - \frac{i}{2 \hbar} \int d^3x \pi_T(\vec x) (\mathrm{h}_{E_A}^T(\vec x)+\mathrm{h}_{E_B}^T(\vec x))\right) \Psi_{vac}[\pi_{ij}] \ket{E}_A \ket{E}_B\ket{\pi_{ij}}_G,
\end{equation}
in which $\mathrm{h}_{E_A}^T(\vec x)$, $\mathrm{h}_{E_B}^T(\vec x)$ are given by Eq.(\ref{deth}). Computing the energy eigenvalue of the gravitational Hamiltonian, we find 
\begin{equation}
{\cal E}_{ABG}   = {\cal E}_{vac} -  \frac{\kappa}{8 \pi}\int {d^3  x}\, d^3  y\ \frac{(E_A(\vec{x})+E_B(\vec{y}))^2}{|\vec{x}-\vec{y}|},
\end{equation}
where ${\cal E}_{vac}$ is the vacuum energy of the field. To identify the relevant energy contribution, we need to subtract from the previous expression the vacuum energy $\cal E$ and the gravitational self-energy of each  source, i.e. \,${\cal E}_{S} = -  \frac{\kappa}{8 \pi}\int {d^3  x}\, d^3  y\ E_{S}(\vec{x}) E_{S}(\vec{x})/|\vec{x}-\vec{y}| $, with $S=A, B$. We  find that the source-dependent contribution to the entangling phase is
\begin{equation}\label{eq:phasewidesource}
\Theta_{AB} = 
-  \frac{\kappa t }{4\pi \hbar}\int {d^3  x}\, d^3  y\  \frac{E_A(\vec{x})E_B(\vec{y})}{|\vec{x}-\vec{y}|},
\end{equation}
which is a generalisation for delocalised quantum sources of the phase of Refs.\,\cite{bose2017spin, marletto2017gravitationally} corresponding to the Newton potential. 

Importantly, we would not have obtained this result had we expanded the general source in a basis of coherent states (i.e.\,those corresponding to semiclassical localised states), and calculated the gravitational field associated to each coherent state. This is the approach taken in Ref.\,\cite{Bengyat:2023hxs}, and it only gives rise to Newtonian interaction. 

We now show that our result cannot be mimicked by some of the most common models in which a full quantum description of the gravitational field is not adopted. The two main possibilities are: i) the gravitational interaction is described via the Newton potential; ii) gravity is kept classical but couples to quantum matter. In the first case, we cannot reproduce the functional expression of the relative phase; in the second case, the model shows nonlinearities or does not generate entanglement between the quantum sources. Hence, we conclude that observing the entanglement phase generated via gravity with two delocalised sources implies that either general relativity needs to be modified \emph{ad hoc} to predict the exact functional form of the phase that we obtain, or gravity needs to have a quantum description.

\subsection{Interaction via the Newton potential}
If we ignore the field character of gravity, the interaction between two source masses  can be expressed via direct coupling through a potential. The standard prescription is to use the Newton potential, and this has been adopted for masses in a quantum superposition of localised states~\cite{bose2017spin, marletto2017gravitationally} as well as for delocalised masses in Gaussian states or harmonic oscillators (see e.g.\,\cite{Qvarfort:2018uag, Liu:2020zyo, Yant:2023smr, Lami:2023gmz, Bengyat:2023hxs, Fujita:2023pia}). We here show that using the Newton potential to describe these scenario corresponds to approximating the quantum source as a superposition of localised sources as in Eq.\,\eqref{eq:suplocstate}. Hence, the Newton potential can be used when the source is in a quantum superposition of semiclassical localised states, but not when the source is in a general quantum state with a large delocalisation. In the latter case, the functional form of the coupling between the two masses obtained using a complete model of gravity in the quantum regime does not coincide with the one obtained using the Newton potential. To see this, let us consider the Hamiltonian describing the interaction between two mass sources 
\begin{equation}
    \hat{H}_N = \frac{\hat{p}^2_A}{2m_A} + \frac{\hat{p}^2_B}{2m_B} + \hat{V}_N (\hat{x}_A - \hat{x}_B),
\end{equation}
where $\hat{V}_N (\hat{x}_A - \hat{x}_B)=- G\frac{m_A m_B}{|\hat{x}_A - \hat{x}_B|}$. The action of the Newton potential on an arbitrary quantum state of the sources $A$ and $B$ is
\begin{equation}\label{Newton}
    \hat{V}_N (\hat{x}_A - \hat{x}_B) \ket{\psi}_A \ket{\phi}_B = -G \int dx_A d x_B \psi(x_A) \phi (x_B) \frac{m_A m_B}{|x_A - x_B|} \ket{x_A}_A \ket{x_B}_B.
\end{equation}
Let us now compare the above expression with the entangling phase which we obtained from the linearised quantum gravity interacting with two  quantum  sources. In particular, we remark that Eq.\,(\ref{eq:phasewidesource})
coincides with the Newton potential in Eq.\,(\ref{Newton}) only in a specific limit. First, the quantum state $\ket{E}_S (S=A,B)$ needs to be approximately a position state $\ket{x}_S$, and more precisely a coherent state $\ket{\alpha_x}_S$ of the type discussed in Sec.\,\ref{sec:SemiclassicalS}. This means that, if the source is a delocalised quantum state, using the Newton potential amounts to approximating the full quantum state of $S$ and $G$ as a linear combination of quantum states as in Eq.\,~\eqref{eq:semiclGrav}. As we have shown, this only gives rise to a quantum superposition of classical gravitational effects. Instead, the phase arising from the full model that we have derived here cannot be reproduced superposing classical Newtonian effects (i.e.\,a classical theory of gravity in the weak-field limit). It is easy to check that the Newtonian phase can only be recovered if $E_S(\vec x) = m_S \delta(\vec x - \vec x_S)$, for $S=A, B$, i.e.\,in the limit in which Eq.\,\eqref{eq:semicl} holds. In future experiments, one could make the effect more explicit by tuning different functional forms of the eigenvalue of $\hat{T}_{00}$, i.e.\,the energy density $E(x)$, and observe how the phase changes with the energy density. This may be achieved, for instance, by using different potentials for the source.

The only possibility of reproducing the phase via a non-local potential, which is however different to the Newton potential, is to introduce an \emph{ad hoc} coupling between the two source masses A and B,  namely
\begin{equation}
    \hat{V}_{Nloc} = - G \int d^3x d^3 y \frac{\text{Tr}[\hat{T}_A (\vec x)] \text{Tr}[\hat{T}_B (\vec y)]}{|\vec x - \vec y|}.
\end{equation}
This coupling is an operator on the matter degrees of freedom of the sources and does not require invoking any gravitational field degree of freedom. However, it would constitute a modification of general relativity, which does not include quantum delocalised sources. In addition, to the best of our knowledge, this coupling is not predicted by any well-known (standard or modified) theory of gravity. 

We now ask if the phase of Eq.\,\eqref{eq:phasewidesource} can be reproduced using a model in which gravity is a classical field and couples to quantum matter. We consider two cases: a genuine classical-quantum coupling such as the one introduced in Ref.\,\cite{oppenheim2023postquantum}, and the Schr{\"o}dinger-Newton equation, which can be seen as the semiclassical limit of the Einstein's equations~\cite{Bahrami:2014gwa}.

\subsection{Hybrid model for classical gravity coupled to quantum matter}

Let us consider the former case. If gravity is classical and is coupled to quantum matter, a general hybrid classical-quantum (CQ) state is~\cite{oppenheim2023postquantum}
\begin{equation}
    \hat{\rho}_{CQ} (t) = \int dz \rho(z, t) \ket{h_{ij}^z}_G \bra{h_{ij}^z} \otimes \hat{\sigma}_S (z, t),
\end{equation}
where $z$ is a variable that correlates the quantum state of the source and the state of the gravitational field. It was shown in Ref.\,\cite{oppenheim2023postquantum} that, if one requires the dynamical law to preserve positivity, the normalisation of probabilities, and the set of CQ states, the most general form of the time evolution is a stochastic open-system dynamics. This type of dynamics is characterised by diffusion and decoherence~\cite{Oppenheim:2022xjr}, and thus would not lead to the generation of an entangled state between the source $S$ and the test particle $P$. Hence, the physical effects due to such a classical-quantum coupling are extremely different to those we obtained by coupling quantum matter to an effective field theory of gravity in the quantum regime, and the two can be distinguished experimentally, for instance by measuring the coherence of the quantum state after the interaction. Notice that such an irreversible coupling is the most general CQ coupling that can be obtained under the reasonable assumptions of Ref.\,\cite{oppenheim2023postquantum}. As shown in Ref.\,\cite{Galley:2023byb} using a theory-independent no-go theorem, if the matter source can be prepared in a quantum superposition state, gravity is classical, and the state of gravity is influenced by the quantum state of matter (back-reaction), then the coupling has to be irreversible. This immediately implies that any genuine coupling between classical gravity and quantum matter has very different observable effects to those discussed in this work.

\subsection{The Schr{\"o}dinger-Newton equation}

Let us now consider the Schr{\"o}dinger-Newton equation. In the Newtonian limit, the interaction Hamiltonian for two particles of mass $m_1$ and $m_2$ interacting gravitationally is~\cite{Giulini:2014rua}
\begin{equation}
    H_I^{SN} \Psi( \vec x_1, \vec x_2, t) = -G \sum_{a=1,2} \sum_{b=1,2} m_a m_b \int d^3x'_1 d^3 x'_2 \frac{|\Psi( \vec x'_1, \vec x'_2, t)|^2}{|\vec x_a - \vec x_b'|} \Psi( \vec x_1, \vec x_2, t),
\end{equation}
where $\Psi( \vec x_1, \vec x_2, t)$ is the joint state of the two particles. In the perturbative regime of gravity, we can simplify the previous expression by taking $\Psi( \vec x_1, \vec x_2, t) = \Psi^{(0)}( \vec x_1, \vec x_2, t) + G \Psi^{(1)}( \vec x_1, \vec x_2, t)$, and the number in brackets refers to the order in perturbation theory in the gravitational constant $G$. In this case, we have $\Psi^{(0)}( \vec x_1, \vec x_2, t) = \psi_1 (\vec x_1) \phi_2 (\vec x_2)$. The leading order of the wave function $\Psi^{(0)}$ is a product state, and the first order of perturbation $\Psi^{(1)}$ characterises the entanglement due to gravity. Hence, we find
\begin{equation} \label{eq:SNphase}
    H_I^{SN} \Psi^{(1)}( \vec x_1, \vec x_2, t) = - G\sum_{a=1,2} \sum_{b=1,2} m_a m_b \int d^3x'_1 d^3 x'_2 \frac{|\psi_1 (\vec x'_1)|^2 | \phi_2 (\vec x'_2)|^2}{|\vec x_a - \vec x_b'|} \Psi^{(0)}( \vec x_1, \vec x_2, t).
\end{equation}
To obtain an expression as close as possible to Eq.\,\eqref{eq:phasewidesource}, we neglect the self-interaction terms in which $a=b$ and we identify $\tilde{E}_1 (\vec x_1) = m_1|\psi_1 (\vec x'_1)|^2$ and $\tilde{E}_2 (\vec x_2)= m_2| \phi_2 (\vec x'_2)|^2$. Notice that this identification is different to the expression of the eigenvalues $E_{1,2} (\vec{x})$ of the $\hat{T}_{00}$ for a quantum source. In this case, Eq.\,\eqref{eq:SNphase} is still not equivalent to Eq.\,\eqref{eq:phasewidesource}. This is not surprising: although the Schr{\"o}dinger-Newton equation can also be obtained as a mean-field approximation of a theory in which gravity is fundamentally quantum~\cite{Bahrami:2014gwa}, this approximation leads to nonlinearities. Our derivation comes from a quantum description of gravity in the weak-field regime and preserves the linearity of quantum theory, hence it is to be expected that the two expressions are different.

In summary, we have shown that both a genuine CQ coupling and the Schr{\"o}dinger-Newton equation lead to quantitatively different predictions to our model, and can hence be distinguished from our result in an experiment.

\section{Quantum commutator of the gravitational field}
\label{sec:commutator}
\begin{figure}[h]
    \centering
\includegraphics[scale=0.25]{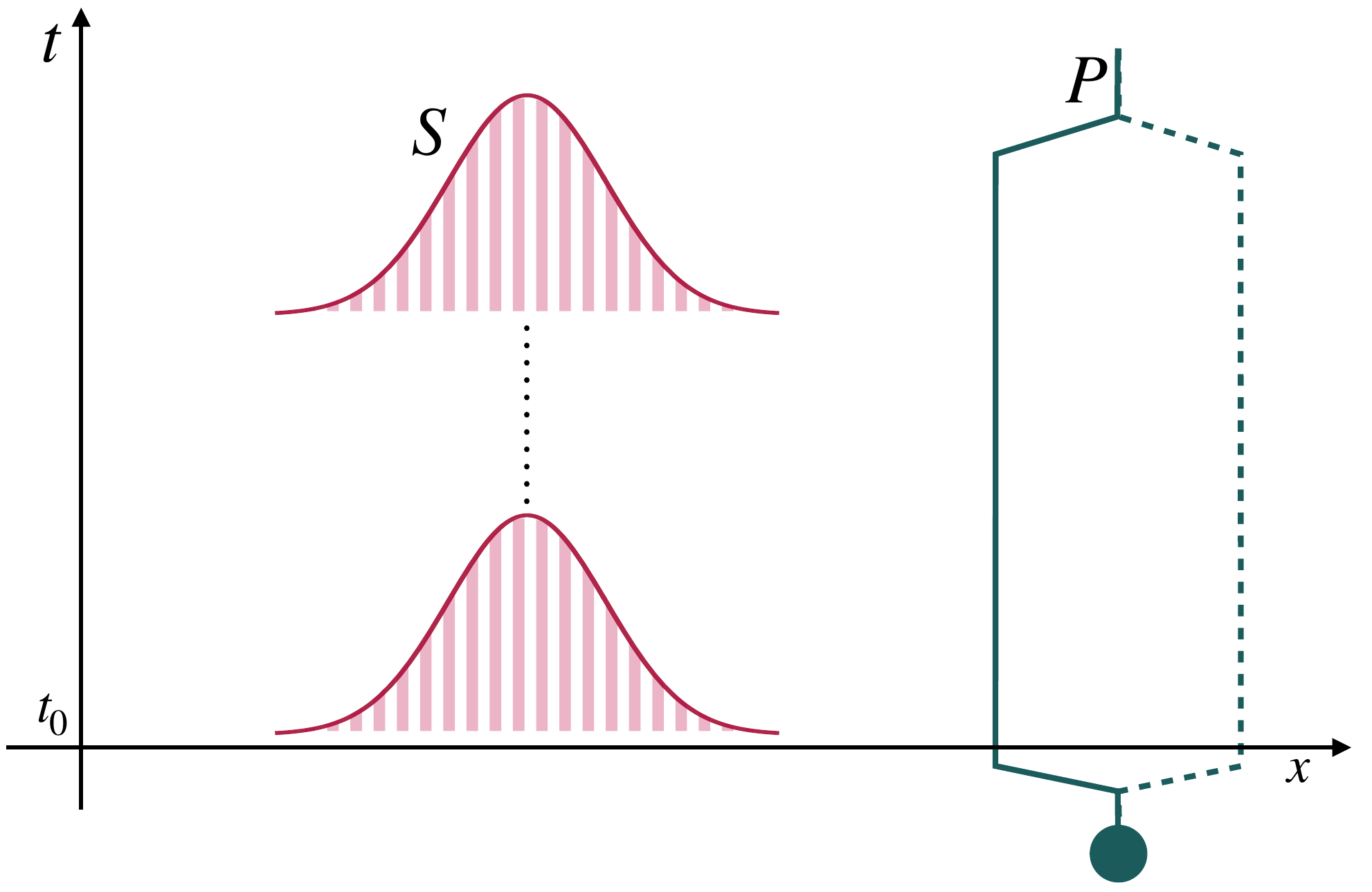}
    \caption{ Before the start of the experiment at $t_0$, a source mass $S$ is prepared in an arbitrary delocalised state, here represented for simplicity as a Gaussian state, and a test mass $P$ (the probe) is initially in a product state with the rest. The source and the moving probe interact gravitationally, and the source is assumed to be static. Using a weak-field quantum description of gravity, after some time the full state of the source, the gravitational field, and the probe becomes entangled. At the end, the source and the probe are measured. The relative phase of the full quantum state encodes the functional form of the gravitational interaction. We find that, for a general quantum state of the source, the gravitational interaction cannot be represented as the Newton interaction and cannot be simulated by a known classical or semiclassical model of gravity. 
   The gravitational commutator appears as an additional term in the relative phase of the quantum source interacting with a test particle.} 
    \label{fig2}
\end{figure}

In this Section, we show that the quantum commutator $[\hat{h}_{ij} (\vec{x}) ,\hat{\pi}^{kl}(\vec{x}')]  =  i \hbar  \delta^{k}_{(i} \delta^{l}_{j)}  \delta^3(\vec{x} - \vec{x}')$ gives rise to additional terms in the relative phase of an interferometric experiment in which two massive objects become entangled via gravitational interaction. We consider the situation depicted in Fig.\,\ref{fig2}, where we have a massive source of gravity $S$ prepared in a general quantum state and a moving test particle $P$, initially in a product state, interacting gravitationally. For more generality, the position of the source $S$ and the momenta of the probe $P$ can be controlled by external potentials. The Hamiltonian is the sum of a free and an interaction term, namely
\begin{equation} \label{eq:fullH}
   \hat{H} = \hat{H}_S + \hat{H}^0_G + \hat{H}_P + \hat{H}_I^{(tot)},
\end{equation}
where $\hat{H}_S$ is the source Hamiltonian, $\hat{H}^0_G$ is the source-free gravitational field Hamiltonian,  $\hat{H}_P$ is the test particle Hamiltonian, and $\hat{H}_I^{(tot)}$ is the interaction Hamiltonian between gravity,  the source $S$ and the probe $P$. In particular, the source Hamiltonian and the test particle Hamiltonian are
\begin{equation}
   \hat{H}_S = \frac{\hat{p}_S^2}{2m_S} + \hat{V}_S(\hat{x}_S), \qquad \hat{H}_P = \frac{\hat{p}_P^2}{2m_P} + \hat{V}_P(\hat{x}_P),
\end{equation}
and $\hat{V}_S(\hat{x}_S)$, $\hat{V}_P(\hat{x}_P)$ are externally-controlled potentials used to control the quantum states of the source and the test particle. The interaction Hamiltonian between gravity and $S$ and $P$ is
\begin{equation}
    \hat{H}_I^{(tot)} = - \frac{1}{2}\int d^3x\, \hat{h}_{\mu\nu}(\vec x) (\hat{T}^{\mu\nu}_S (\vec x)+ \hat{T}^{\mu\nu}_P (\vec x)).
\end{equation}
Working in the temporal gauge $h_{0\mu} =0$, the total Hamiltonian of the source, the test particle, and the gravitational field can be equivalently written as
$
   \hat{H} = \hat{H}_S + \hat{H}_G + \hat{H}_P + \hat{H}_I,
$
where $\hat{H}_G$ is the Hamiltonian of Eq.\,\eqref{eq:Hgr} with the same constraints as in Eq.\,\eqref{eq:constwide}\footnote{Notice that, although adding another particle would in principle change the constraints, our assumption that the test particle does not back-react on the gravitational field implies that the quantum state of gravity of the source alone stays within the constraint surface.} and $\hat{H}_I$ is the interaction Hamiltonian
\begin{equation} \label{eq:intH}
    \hat{H}_I = -\frac{1}{2} \int d^3x \hat{h}^{ij} (x)\hat{T}^P_{ij}(x).
\end{equation}
In the specific case we consider, we make the approximation that the test particle does not back-react on the gravitational field.  In other words, we ignore the gravitational field of the probe, and only consider its interaction with the gravitational field that the source generates. More concretely, this means that the interaction with the probe does not change the quantum state of the gravitational field sourced by $S$ alone nor the expectation value of the gravitational field operators. This approximation ensures that adding the interaction Hamiltonian between the gravitational field and the test particle does not map the the quantum state of the gravitational field out of the constraint surface, and hence that the eigenstate of the initial Hamiltonian remains the same.  We take as the initial state
\begin{equation}
|\Xi_0\rangle_{SGP} = |\Psi_0\rangle_{S+G}\otimes  |\phi_0\rangle_P,
\end{equation}
where $|\Xi_0\rangle_{S+G}$ and $|\phi_0\rangle_P$ are arbitrary quantum states of the source and of the probe respectively. 

The final state at time $t$ is 
\begin{equation}\label{eq:evolutionU}
|\Xi_t\rangle_{SGP} = \hat{U}_{SGP}(t) |\Xi_0\rangle_{SGP} = e^{- \frac{it}{\hbar} \hat{H}} |\Xi_0\rangle_{SGP}.
\end{equation}
Since the test particle does not back-react on the gravitational field,  the dynamical evolution preserves the constraints, namely the initial physical state is still a solution of the constraint equation after the time evolution. Technically, this means that the interaction Hamiltonian weakly commutes with the constraints, i.e.
\begin{equation}
    \left[\hat{H}_I, \hat{\mathcal{C}}^i\right] \ket{\Xi}_{S+G} = - \frac{1}{2} \int d^3k \, k'_j 
    \hat{T}_{lm}^P (-\vec k) \left[\hat{h}^{lm}(\vec k), \hat{\pi}_L^{ij}(\vec k')\right]\ket{\Xi}_{S+G} =0.
\end{equation}
This is equivalent to neglecting the commutator between the gravitational field operator and the longitudinal momentum operator when it acts on the physical state, i.e.\,$\left[\hat{h}^{lm}(\vec k), \hat{\pi}_L^{ij}(\vec k')\right]\ket{\Xi}_{S+G} =0$.
Conversely, the commutator of the transverse mode $\left[\hat{h}^{lm}(\vec k), \hat{\pi}_T^{ij}(\vec k')\right]\ket{\Xi}_{S+G}$ cannot be neglected, because it is responsible for the physical effects.
The crucial observation is that the free Hamiltonian of the gravitational field $\hat{H}_G$ and the interaction Hamiltonian $\hat{H}_{I}$ do not commute when applied to the physical state. This can be seen explicitly (see details in App.~\ref{app:commutator}), namely
\begin{equation}
 [\hat{H}_G ,\hat{H}_I]\ket{\Xi}_{S+G} = i \hbar \kappa  \int d^3 x \left( \hat{\pi}_{ij}^T(\vec{x})  - \frac{1}{2}P_{ij} \hat{\pi}^T(\vec{x}) \right)   \hat{T}^{ij}_P (\vec{x})\ket{\Xi}_{S+G}.
\end{equation} 
In the following, we focus only on the terms  arising from the commutator between $\hat{H}_G$ and $\hat{H}_I$, and we neglect the rest of the commutators (for instance, the commutators between the position and momentum of $S$ and $P$), which are not relevant for the effect we wish to describe. These terms give rise to an additional phase that can be distinguished experimentally, thanks to its different functional form, from the gravitational commutator terms. In addition, the commutators between $\hat{H}_G$ and $\hat{H}_I$ give rise to a polynomial expression in the time of the experiment $t$\footnote{Note that the evolution time $t$   corresponds to the time measured in frame of  temporal gauge, in which $g_{00}$ is fixed to be 1. In our laboratory frame, however, the corresponding time $t_{lab}$ of the evolution is actually shorter. This difference  is negligible for the physical effect we are considering, since the  correction is at the order of $G$:  $ t_{lab}/t  \sim 1-\mathcal{O} (\frac{m_s G}{r}) $ , in which $r$ is the relative distance between the source and the probe.  }. We here evaluate terms up to order $t^3$. Considering higher order terms is more challenging from a computational perspective, but has no impact on the final result, namely the dependence of the relative phase on the commutator between the gravitational field operator and its canonically conjugated momentum. The full calculation is detailed in App.\,\ref{app:commutator}.

We expand the initial quantum state in its energy eigenbasis as
\begin{equation}
|\Xi_0\rangle_{SGP} = \int d\mu(E_S)   \psi_S(E) \ket{E}_S \ket{g_E}_G \ket{\psi}_P 
\end{equation}
where $\ket{g_E}_G$ was defined in Eqs.\,(\ref{scalarcons}\,-\ref{statepi}). The state at time $t$ is
\begin{equation}
|\Xi_t\rangle_{SGP} = \int d\mu(E) \psi_S(E)  e^{-\frac{it}{\hbar}\theta_{SP}} e^{\frac{i}{\hbar}(\hat{\Theta}^{(0)} + \hat{\Theta}^{(1)} + \hat{\Theta}^{(2)}+\cdots)}\ket{E}_S \ket{g_E}_G \ket{\psi}_P,
\end{equation}
in which $\theta_{SP}$ is a phase obtained from the action of the source and test particle Hamiltonian on the quantum state, and is irrelevant to our purposes, and the index $n=0, 1, 2$ in each $\hat{\Theta}^{(n)}$ is the number of commutators between the gravitational field operators that give rise to the phase. 
More explicitly,
\begin{equation}
    \hat{\Theta}^{(0)} = - \frac{ t}{4 } \int \frac{dk^3}{(2 \pi)^3} \textnormal{h}_{E}^T(\vec k)P_{ij} \hat{T}^{ij}_P(-\vec k)  -\frac{i \kappa t^2}{8 } \int \frac{d^3 k}{(2 \pi)^3} \frac{1}{|k|} \left( \hat{T}^P_{ij}(\vec k)\hat{T}_{P}^{ij}(-\vec k) - 2 \hat{\tilde{T}}^{ij}_{PT}(\vec k) \hat{\tilde{T}}_{ij}^{PT} (-\vec k) \right) .
\end{equation}
The first term in the previous expression is an additional part of the entangling phase coming from the coupling between the gravitational field and the momentum of the probe. The second term, which is not a phase but a real exponential term, comes from the integration of the transverse traceless component of the gravitational field, after evaluating the operators on its quantum state. 

The term $\hat{\Theta}^{(1)}$ corresponds to the phase arising from calculating a single commutator of the gravitational field operators, i.e.\,$[\hat{H}_G ,\hat{H}_I]$, namely
\begin{equation}
   \hat{\Theta}^{(1)} = \frac{   \kappa t^3}{8 } \int \frac{d^3 k}{(2 \pi)^3}  \hat{T}^P_{ij}(\vec k)\hat{T}_{P}^{ij}(-\vec k) .
\end{equation}
The term $\hat{\Theta}^{(2)}$ is instead the phase arising due to the double commutators $[\hat{H}_G, [\hat{H}_G ,\hat{H}_I]]$ and $[\hat{H}_I, [\hat{H}_G ,\hat{H}_I]]$, specifically
\begin{equation} \label{eq:Theta2}
    \begin{split}
        \hat{\Theta}^{(2)} &= \frac{ \kappa t^3}{6}  \int \frac{dk^3}{(2 \pi)^3} \left[\hat{T}_P^{ij} (\vec k)  P^k_i P^l_j \hat{T}^P_{kl} (-\vec k) - \frac{1}{2}  (P_{ij} \hat{T}_P^{ij}(\vec k)) (P_{kl} \hat{T}_P^{kl}(-\vec k)) \right]\\
    &=\frac{ \kappa t^3}{6} \int \frac{d^3k}{(2 \pi)^3}  \hat{\tilde{T}}^{PT}_{ij}(\vec k)   \hat{\tilde{T}}_{PT}^{ij}(-\vec k).
    \end{split}
\end{equation}
Note that the $\hat{\Theta}^{(n)}$ operators, with $n=0,\,1,\,2$, do not depend on the gravity degrees of freedom, but only on  the test particle $P$. In the second line, $\tilde{T}_{PT}^{ij}$ denotes the transverse traceless component of $T_{P}^{ij}$. In addition, all terms in the phase are of the order of $\frac{\kappa}{\hbar}$. This is natural, because we are expanding the gravitational field to the first order in perturbation theory. This means that we are not considering higher order terms giving rise to graviton loops corresponding to the self-interaction of the gravitational field with itself.  Hence, the order in the gravitational coupling and $\hbar$ is the same as in the GIE proposals. In addition, $ \hat{\Theta}^{(1)}$ and $ \hat{\Theta}^{(2)}$ scale differently with time to the leading order phase that does not depend on the commutator, hence this term can in principle be distinguishable experimentally. 
Therefore, the $\hat{\Theta}^{(n)}$ operators, with $n=0,\,1,\,2$, give rise to a relative phase which appears in the interference pattern between different  eigenstates of the probe.

To observe this relative phase, the probe needs to be prepared in a quantum superposition of states on which the action of the stress energy tensor gives different values. This can be realised, for instance, when the probe is prepared in a quantum superposition of different momentum or energy eigenstates (possibly with a large energy gap). It is important to note that, after the interaction with the gravitational field, the full quantum state of the probe, the source, and the gravitational field is entangled. Hence, a joint measurement on the source and the probe, in which the gravitational field follows the quantum state of the source adiabatically (this condition is equivalent to requiring no emission of gravitational radiation)~\cite{Unruh:1999vn}, should be performed in order to preserve the coherence of the full state.

\section{Discussion}

In this work, we identify two physical effects to test the quantum nature of the gravitational field in table-top experiments, which cannot be predicted using the Newton potential.
The analysis is carried out using  a quantum field formulation of gravity in the weak-field regime. Instead of the standard Fock space representation, we use the Schr{\"o}dinger representation in the field basis. This formulation is not only convenient to identify quantum states of macroscopically distinct gravitational configurations; most importantly, it is suitable to describe the quantum state of  the gravitational field given by any generic massive quantum source. Hence, it provides a framework to investigate rich physical properties and predictions that will be explored in future quantum test of gravity in table-top experiments.

The first physical effect that we study is encoded in the interference pattern of two quantum sources of gravity in delocalised states. We show that, when the massive sources are in a generic quantum state, the dynamical phase arising from their interaction  cannot be reproduced by the Newton potential nor by any theory in which gravity is classical. Therefore, observing the functional form of such a phase in a future table-top experiment would imply that gravity needs to have a quantum description (or an unknown description which, to the best of our knowledge, cannot be reproduced by any known theory of gravity). In the limit in which the source is in a superposition of semiclassical localised states, as in the case in proposals to observe gravitationally-induced entanglement (see e.g.\,\cite{bose2017spin, marletto2017gravitationally}), the prediction reduces to the one given by the Newton potential. Hence, all our results are consistent with those studied in the literature, when reduced to the same regime. Importantly, this effect is of the same order in the gravitational field strength as the Newton potential.

The second physical effect that we study is how the quantum commutator of the gravitational field enters the expression of the dynamical phase of an interferometric configuration involving a source particle and a test particle. Specifically, we show that in the entangling phase between a moving probe and a mass source, the commutators between the linearised gravitational field operators and its conjugate momenta enter as additional terms. Probing these terms would be a substantial advantage compared to Newton entanglement, because it would provide an explicit signature that the gravitational field is a quantum mediator. Both effects we identified scale the same  as the Newton potential in terms of the gravitational constant, however they are more challenging to measure experimentally. Future work will investigate how these effects, identified here for the first time, can concretely be observed, and the techniques to overcome the challenges.

It's worth noting that a similar analysis can be conducted for the electromagnetic field, mirroring the methodology in this paper but technically much simpler. The electromagnetic version of these effects may be in the reach of experiments and could serve as a compelling proof-of-concept demonstration for the physical effects identified in this paper.  Another possibility for a proof-of-principle experiment is to use an optomechanical systems with phonon propagation, e.g.\,along the lines of Ref.\,\cite{Xu:2021zej}. The study of a concrete experimental implementation of these ideas is left for future work.

In conclusion, we have shown for the first time that, even in the case of static quantum sources of gravity, there are gravitational effects, observable in future table-top experiments, that give stronger evidence of the quantum nature of gravity than those only requiring the Newton potential. Specifically, such effects cannot be explained using the Newton potential nor, to the best of our knowledge, any known model in which gravity is classical. Identifying these effects is crucial from a conceptual perspective, because they increase the impact of current efforts to test the quantum nature of gravity. Identifying and eventually realising alternative experiments testing stronger and diverse quantum effects in the gravitational field would open for the first time an observational window on quantum effects in gravity.

\acknowledgments{We are very grateful to Onur Hosten for providing substantial comments for the second version of the manuscript. We would like to acknowledge Miles P.\,Blencowe, Carlo Rovelli and Renato Renner for useful feedback on a first draft of this work. We would also like to thank Markus Aspelmeyer, Francesco Becattini, Yiwen Chu, Tao Shi, and Julian Sonner for helpful discussions and comments. F.G. and L.Q.C. acknowledge support from the ETH Zurich Quantum Center, NCCR SwissMAP, and from the John Templeton Foundation, as part of the \href{https://www.templeton.org/grant/the-quantuminformation-structure-ofspacetime-qiss-second-phase}{‘The Quantum Information Structure of Spacetime, Second Phase (QISS 2)’ Project}. F.G. acknowledges support from the Swiss National Science Foundation via the Ambizione Grant PZ00P2-208885. The opinions expressed in this publication are those of the authors and do not necessarily reflect the views of the John Templeton Foundation.}

\appendix

\section{The quantum state of the gravitational field for a localised semiclassical source}
\label{app:SemClassG}

We here review the quantisation procedure of the weak gravitational field of Ref.\,\cite{Chen:2022wro}. Table-top experiments for quantum tests of gravity can be described in the regime of a weak gravitational field with non-relativistic matter sources.  In order to provide a suitable theoretical framework  for those experiments, we quantise linearised gravity in the field basis~\cite{Hatfield:1992rz}.  To the first order of the perturbations around the flat metric, i.e.\,$g_{\mu\nu} = \eta_{\mu\nu} +  h_{\mu\nu}$, the classical action of linearised gravity~\cite{Carroll:2004st} coupled to a matter field reads
\begin{equation}\label{linearzedgravityaction}
	S = \frac{1}{4\kappa} \int  d^4x  \left(-	\partial_{\mu} {h}_{\alpha \beta} \partial^{\mu} {h}^{\alpha \beta} + 	\partial_\mu {h}  \partial^\mu {h} 	-2 \partial_{\mu} {h}^{\mu \nu} \partial_\nu {h} +2 \partial_\alpha {h}_{\mu \nu} \partial^\mu {h}^{\alpha\nu} \right) + \frac{1}{2}\int d^4x\, {h}_{\mu \nu} T^{\mu \nu} 
   \end{equation}
	in which $\kappa = 16 \pi G/c^4$.  For canonical quantisation, the metric is cast in the 3+1 decomposition: $ds^2 = - N^2 dt^2 + \gamma_{ij} (dx^i + N^i dt) (dx^j + N^j dt)$, in which  $\gamma_{ij}$ is the metric of the space-like foliation. $N$ and $N^i $  are the lapse and  the shift vector respectively; the lapse is a gauge parameter corresponding to the rate of time flow between different foliations, and the components of the shift vector are the gauge parameters relating the spatial coordinates on different foliations at  different times. The perturbation in the 3+1 decomposition is expressed as
	$\gamma_{ij} = \delta_{ij} +  h_{ij},\  N = 1 +n,\ N^i =0+n^i $, therefore $h_{00}  =  - 2n, h_{0j} = \delta_{ij} n^i = n_j, h^{0i}  = - n^i.$ From the action, one can derive the canonical  momenta:  $ \pi^{kl} = \frac{1}{2\kappa} ( \dot{h}^{kl} - \dot{h} \delta^{kl} ) $. The canonical momenta and the linearised gravitational field satisfy the Poisson brackets: $\{ h_{ij} (\vec{x}), \pi^{kl} (\vec{x'})\} = \delta^k_{(i} \delta^l_{j)} \delta^3(\vec{x}-\vec{x}') $.

	An important feature of our quantisation procedure is that we fix the gauge minimally.  We only perform the partial gauge fixing $\partial_i n_j =0$ and  $\partial_i n =0$ and keep spatial diffeomorphism invariance. We obtain the Hamiltonian:
	\begin{equation} \label{FullHami}
		H_{G+S}  = \kappa  \int d^3x  \  \left(\pi_{kl} \pi^{kl} - \pi^2/2 \right) + \frac{1}{4\kappa}   \int d^3x [ \partial_k  h_{ij} \partial^k  h^{ij}  - \partial_i h \partial^i h - 2 \partial^k h_{ik} (\partial_j h^{ij}-\partial^i h) +\frac{1}{2} h_{ij}T^{ij}- 4 n_i  \mathcal{G}^i - 4n \mathcal{C} ], 
\end{equation}
where the perturbation of the lapse $n$ and the shift vector $n_i$ become Lagrangian multipliers imposing four commuting constraints.
 In particular, the vector constraints ${\cal G}^i$ and scalar constraint $	\mathcal{C}_{\rho} $ are
 \begin{equation}
{\cal G}^i=2 \partial_j \pi^{ij} +  T^{0i} =0, \ \ 
	\mathcal{C}_{\rho}  = - \partial_i \partial^i  h^T -\kappa T^{00} =0.
\end{equation}
Since the matter source is  quasi-static in the laboratory frame, $T^{0i}$ and $T^{ij}$ are negligible compared to  $T^{00}$. Therefore, the  information on the matter source is encoded in the gravitational field only through the scalar constraint $\mathcal{C}_{\rho}$.
If the source $S$ is in a quantum state $| \alpha_i  \rangle_S $, i.e.\,a semiclassical (i.e.\,no position-momentum commutator effect is measurable on the quantum state up to experimental resolution) and localised quantum state, it is approximately and eigenstate of $\hat{T}_{00}$.  In this case, we have  (see Eq.\,\eqref{eq:semicl})
\begin{equation}
	\hat{T}_{00} | \alpha_i  \rangle_S   \approx  \rho_i (x, t)| \alpha_i  \rangle_S \approx  m c^2  \delta^3(\vec{x}-\vec{x}_i) | \alpha_i  \rangle_S,
\end{equation}
which describes a point-like particle with rest mass $m$. The trace of the transverse components of the metric perturbation, $h_T$, is obtained by solving the Poisson equation through the scalar constraint $\mathcal{C}_{\rho}$:
\begin{equation}
\mathsf{h}^T_{\rho_i}(\vec {x}) =
\frac{\kappa}{4\pi}\int d^3 y \frac{\rho_i(\vec{y})}{ |\vec{x} - \vec{y} |} + f(\vec {x})  = \frac{\kappa}{4\pi} \frac{m c^2}{ |\vec{x} - \vec{x_i} |} + f(\vec {x}).
\end{equation}
Here,  $f(\vec{x}) $ can be an arbitrary harmonic function and it is fixed by the spacetime boundary conditions. 

For a quasi-static source, the full Hamiltonian of the gravitational field interacting with the source in Eq.\,\eqref{FullHami} becomes source-free as in  Eq.\,\eqref{eq:Hgr}, namely
\begin{equation}
\hat{H}_G = \kappa  \int \frac{d^3k}{(2 \pi)^3}  \  \left(\pi_{kl} \pi^{kl} - \pi^2/2 \right) + \frac{1}{4\kappa}   \int \frac{d^3k}{(2 \pi)^3} \left( k^2  h^T_{ij}(\vec{k})  h_T^{ij}(-\vec{k})  - k^2 h_T(\vec{k}) h_T(-\vec{k})\right).
\end{equation}
This Hamiltonian can be quantised using the Dirac prescription for the linearised gravitational field (for further details, see Ref.\,\cite{Chen:2022wro}). For a static (semiclassical and localised) source, the quantum state of the gravitational field is the ground state of the gravity Hamiltonian $\hat{H}_G$, and additionally has to satisfy the scalar and vector constraints. The quantum state of the source and its gravitational field, in both $h_{ij}$ and $\pi_{ij}$ representations, reads
\begin{equation}\begin{split}
|\Psi\rangle_{S+G} &=\eta  \int \mathcal{D}[h_{ij}] \delta[h^T-\mathsf{h}^T_{\rho_{\alpha } } ] \exp\left\{ - \frac{1}{4 \kappa \hbar} \int \frac{d^3k}{(2\pi)^3}\  |\vec{k}|\    h^T_{ij}(\vec{k})  h_T^{ij} (-\vec{k}) \right\} \ket{\alpha }_S \ket{h_{ij}}_G \\
&= \eta'  \int \mathcal{D}[\pi_{ij}] \exp \left\{ - \frac{i}{2 \hbar} \int \frac{d^3k}{(2 \pi)^3} \pi_T(\vec k)  \mathrm{h}^T_{\rho}(\vec k) - \frac{\kappa} {\hbar}\int \frac{d^3k}{(2 \pi)^3} \frac{1}{|k|} \left(   \pi^T_{ij}(\vec k)   \pi_T^{ij}(-\vec k) - \frac{1}{2} \pi_T(\vec k)  \pi_T(-\vec k)  \right) \right\} \ket{\alpha }_S \ket{\pi_{ij}}_G.  
\end{split}
\end{equation}
In a more compact notation, since the source is in an eigenstate of $\hat{T}_{00}$, we can denote the full state as $|\Psi\rangle_{S+G}:=\ket{\alpha }_S \ket{g_{\alpha}}_G$.
If one superposes different quantum states of such matter sources, each of which is centred around a different position $x_i$, i.e.\,$\ket{\psi}_S = \sum_i c_i \ket{\alpha_i}_S$, we can solve the Gauss constraint locally for each quantum state $\ket{\alpha_i}$. The resulting solution is a superposition of classical spacetimes, formally expressed as
\begin{equation}
|\Psi\rangle_{S+G}:=\sum_i c_i \ket{\alpha_i}_S \ket{g_{\alpha_i}}_G.
\end{equation}
This is exactly the description for the matter source and the gravitational field considered in the GIE proposals. The states of the source correspond to semiclassical configurations of matter where the position is localised around a value $x_i$ and the momentum is localised around $p=0$.

 Let us consider two such localised semiclassical quantum states of the source $S$, $\ket{\alpha_{x}}_S$ and $\ket{\alpha_{x+\epsilon}}_S$, which are related by a translation and whose central positions differ by an arbitrarily small amount $\epsilon$. The solutions of the Poisson equation, respectively $\mathrm{h}^T_{\alpha_{x}}(\vec x) $ and $\mathrm{h}^T_{\alpha_{x+\epsilon}}(\vec x)$, are also related by a translation. In particular, the representation of the quantum state in momentum space only differs by a phase  $\mathrm{h}^T_{\alpha_{x}}(\vec k) = e^{-ik\epsilon}\mathrm{h}^T_{\alpha_{x+\epsilon}}(\vec k)$. Such a small difference leads to a vanishing scalar product between the respective full quantum states of matter and gravity, i.e.
 \begin{equation} \label{eq:appscalarprsemiclassical}
 \begin{split}
 \braket{\Psi_{x} | \Phi_{x+\epsilon}}_{S+G}& = \eta^2 \int \mathcal{D} [\pi^T]  \mathcal{D} [\tilde{\pi}^T_{ij}] \mathcal{D} [\pi^L_{ij}] e^{ - \frac{i}{2 \hbar} \int \frac{d^3k}{(2 \pi)^3} \pi_T(\vec k)  \left(\mathrm{h}^T_{\alpha_{x}}(\vec k) - \mathrm{h}^T_{\alpha_{x+\epsilon}}(\vec k)  \right) } e^{ - \frac{2 \kappa} { \hbar } \int \frac{d^3k}{(2 \pi)^3} \frac{1}{|\vec{k}|}  \tilde{ \pi}^T_{ij}(\vec k)   \tilde{\pi}_T^{ij}(-\vec k) } \braket{\alpha_{x}|\alpha_{x+\epsilon}}\\
 & = \delta (\mathrm{h}^T_{\alpha_{x}}(\vec k)- \mathrm{h}^T_{\alpha_{x+\epsilon}}(\vec k) ) \braket{\alpha_{x}|\alpha_{x+\epsilon}} =0.
  \end{split}
  \end{equation} 
The quantum states of the sources, however, have a non-negligible overlap, i.e.\,$\braket{\alpha_{x}|\alpha_{x+\epsilon}} \neq 0$, hence we would expect the scalar product of the source and its gravitational field to be non-vanishing. The vanishing of Eq.\,\eqref{eq:appscalarprsemiclassical} means that the approximation $	\hat{T}_{00} | \alpha_i  \rangle_S    \approx  \rho_{cl}(\vec x) | \alpha_i  \rangle_S$, with $\rho_{cl}(\vec x)$ being the classical energy density, only holds for matter sources prepared in perfectly distinguishable states. Crucially, the vanishing of the scalar product only depends on approximating the eigenvalue of $\hat{T}_{00}$ with the classical function, but not on the specific form of the function. For technical simplicity, we have used  $\rho_{cl}(\vec x) = m c^2  \delta^3(\vec{x}-\vec{x}_i)$, but the scalar product would have also vanished had we used a more realistic energy density, for instance $\rho_{cl}(\vec x) = \frac{m c^2}{\sigma}  e^{-\frac{(\vec{x}-\vec{x}_i)^2}{2 \sigma^2}}$.

\section{Details of the calculations of the scalar product}
\label{app:DetQuantumSource}
Given two  generic quantum  source  $|\psi \rangle_S$ and $|\varphi\rangle_S$, 
the inner product of the full quantum state of the gravitational field together with the source is
\begin{equation}\begin{split}
\langle \Psi_{\psi}|\Psi_{\phi} \rangle_{S+G} & = \eta^2 \int  dx  d\mu(\pi_{ij})d\mu(E) d\mu(E')   \psi_E \phi_{E'}^*  \langle E' | x \rangle \langle x | E \rangle  e^{ - \frac{i}{2 \hbar} \int \frac{d^3k}{(2\pi)^3} \pi_T(\vec k) (\mathrm{h}_{E}^T(\vec k)-\mathrm{h}_{E'}^T(\vec k))} |\Psi_{vac}[\pi_{ij}] |^2 \\
& = \eta^2 \int   d\mu(\pi_T)d\mu(E) d\mu(E')   \psi_E \phi_{E'}^*  \langle E' | E \rangle  e^{ - \frac{i}{2 \hbar} \int \frac{d^3k}{(2\pi)^3} \pi_T(\vec k) (\mathrm{h}_{E}^T(\vec k)-\mathrm{h}_{E'}^T(\vec k))} \int d\mu(\tilde{\pi}^T_{ij})   d\mu(\pi^L_{ij}) |\Psi_{vac}[\pi_{ij}] |^2 \\
& = \int d \mu(E) \psi_E \phi^*_E  = \langle \psi| \phi \rangle 
\end{split}
\end{equation}

\section{Details of the calculations of the quantum commutator}
\label{app:commutator}
In this section, we detail the steps to obtain the relative phase in Sec.\,\ref{sec:commutator} due to the commutator of the gravitational field operators \begin{equation}
[\hat{h}_{ij} (\vec{k}) ,\hat{\pi}^{kl}(\vec{p}')]  =  i \hbar  \delta^{k}_{(i} \delta^{l}_{j)}  \delta^3(\vec{k} + \vec{p}')
\end{equation}
The Hamiltonian of the gravitational field is in Eq.\,(\ref{eq:Hgr})
\begin{equation}
\hat{H}_G = \kappa  \int \frac{d^3k}{(2 \pi)^3}  \  \left(\hat{\pi}_{kl}(\vec{k}) \hat{\pi}^{kl}(-\vec{k}) - \hat{\pi}^2(\vec{k})/2 \right) + \frac{1}{4\kappa}   \int \frac{d^3k}{(2 \pi)^3} \left( k^2  \hat{h}^T_{ij}(\vec{k})  \hat{h}_T^{ij}(-\vec{k})  - k^2 \hat{h}_T(\vec{k}) \hat{h}_T(-\vec{k})\right),
\end{equation}
and the interaction is described in Eq.\,(\ref{eq:intH}) 
\begin{equation} 
    \hat{H}_I = -\frac{1}{2} \int \frac{d^3k}{(2 \pi)^3} \hat{h}^{ij} (\vec{k})\hat{T}^P_{ij}(-\vec{k}).
\end{equation}
The time evolution operator in Eq.\,(\ref{eq:evolutionU}) $\hat{U}_{SGP}(t) : = e^{-\frac{it}{\hbar}\hat{H}} = e^{-\frac{it}{\hbar}(\hat{H}_S + \hat{H}_P + \hat{H}_G+  \hat{H}_{I})} $ can be expanded using the Zassenhaus formula. Specifically, for two general quantum operators $\hat{A}$ and $\hat{B}$, we have 
\begin{equation}
e^{\hat{A} + \hat{B}}= e^{\hat{A} }e^{ \hat{B}} e^{-\frac{1}{2} [\hat{A}, \hat{B}]} e^{ \frac{1}{6} \left([\hat{A},[\hat{A}, \hat{B}] ]+  2[\hat{B}, [\hat{A}, \hat{B}]] \right) }  e^{ -\frac{1}{24} \left([[[\hat{A}, \hat{B}],\hat{A}],\hat{A} ]+  3 [[[\hat{A}, \hat{B}],\hat{A}],\hat{B} ]+  3 [[[\hat{A}, \hat{B}],\hat{B}],\hat{B} ]\right) } ...
\end{equation}

The relevant terms for the physical effect we are studying are those coming from the  gravitational commutator. Hence, we do not explicitly calculate here the terms coming from the commutators $[\hat{x}, \hat{p}]$ of the source and the probe, contained in $\hat{H}_S$ and $\hat{H}_P$.
Expanding the time evolution operator through the Zassenhaus formula, we obtain
\begin{equation} \label{UZasDec}
\hat{U}_{SGP}(t) = e^{-\frac{it}{\hbar}( \hat{\mathcal{H}}_S +\hat{\mathcal{H}}_P)}  e^{- \frac{it}{\hbar}\hat{\mathcal{H}}_G }  e^{- \frac{it}{\hbar} \hat{ \mathcal{H}}_I } e^{\frac{t^2}{2 \hbar^2}[ \hat{\mathcal{H}}_G , \hat{\mathcal{H}}_I ]}
e^{
 \frac{i t^3}{6 \hbar^3} \left([ \hat{\mathcal{H}}_G,[ \hat{\mathcal{H}}_G , \hat{\mathcal{H}}_I ] ]+ 2[ \hat{\mathcal{H}}_I , [ \hat{\mathcal{H}}_G , \hat{\mathcal{H}}_I ]]  \right) } 
 e^{-
 \frac{ t^4}{24 \hbar^4} [ [ [ \hat{\mathcal{H}}_G , \hat{\mathcal{H}}_I ], \hat{\mathcal{H}}_G ],\hat{\mathcal{H}}_G ] }...
\end{equation}
We assume that  the probe as a test particle does not back-react to the gravitational field. Therefore, the interaction with the probe does not take the quantum state of gravitational field out of the constraint surface. Formally, this means that the commutator between the interaction Hamiltonian and the vector constraint is effectively negligible
\begin{equation}
\left[\hat{h}^{lm}(\vec k), k'_i\hat{\pi}^{ij}(\vec k')\right]\ket{\Psi}_{S+G} =\left[\hat{h}^{lm}(\vec k), \hat{\pi}_L^{ij}(\vec k')\right]\ket{\Psi}_{S+G} =0.
\end{equation}
Together with the condition that the quantum gravitational state satisfies the constraint $\hat{\mathcal{G}}_i \ket{\Psi}_{S+G} =0$,  we have
\begin{equation}
\begin{split}
 [\hat{H}_G ,\hat{H}_I]\ket{\Psi}_{S+G} & =\left[\kappa  \int \frac{d^3k}{(2 \pi)^3}  \  \left(\hat{\pi}_{kl}(\vec{k}) \hat{\pi}^{kl}(-\vec{k}) - \frac{1}{2}\hat{\pi}(\vec{k})\hat{\pi}(-\vec{k})\right), -\frac{1}{2} \int \frac{d^3k'}{(2 \pi)^3} \hat{h}^{ij} (\vec{k}')\hat{T}^P_{ij}(-\vec{k}') \right]\ket{\Psi}_{S+G}\\
 & = i \hbar \kappa  \int \frac{d^3k}{(2 \pi)^3}\left( \hat{\pi}_{ij}^T(\vec{k})  - \frac{1}{2}P_{ij} \hat{\pi}^T(\vec{k}) \right)   \hat{T}^{ij}_P (-\vec{k})\ket{\Psi}_{S+G}\\
  & = i \hbar \kappa  \int \frac{d^3k}{(2 \pi)^3} \hat{\tilde{\pi}}_{ij}^T(\vec{k}) \hat{T}^{ij}_P (-\vec{k})\ket{\Psi}_{S+G},
 \end{split}
\end{equation} 
where $\tilde{\pi}_{ij}^T(\vec{k})$ is the transverse traceless part of the momentum of the gravitational field. There are two double commutators: $[ \hat{H}_G,[ \hat{H}_G , \hat{H}_I] ]$ and $ [ \hat{H}_I ,[ \hat{H}_G , \hat{H}_I ]] $. When they act on the physical state, we obtain 
 \begin{equation} \begin{split}
[ \hat{H}_G,[ \hat{H}_G , \hat{H}_I] ] \ket{\Psi}_{S+G} &= \frac{i \hbar }{4}    \left[ \int \frac{d^3k}{(2 \pi)^3}k^2  \left(   \hat{h}^T_{ij}(\vec{k})  \hat{h}_T^{ij}(-\vec{k})  -  \hat{h}_T(\vec{k}) \hat{h}_T(-\vec{k})\right),   \int \frac{d^3k'}{(2 \pi)^3} \left( \hat{\pi}_{kl}^T(\vec{k}')  - \frac{P_{kl}}{2} \hat{\pi}^T (\vec{k}') \right)   \hat{T}^{kl}_P  (-\vec{k}')\right] \ket{\Psi}_{S+G}\\
&= -\frac{ \hbar^2 }{4}   \int \frac{d^3k}{(2 \pi)^3}k^2   \left(2  \hat{h}^T_{ij}(\vec{k})  -  P_{ij} \hat{h}^T(\vec{k}) \right) \hat{T}^{ij}_P(-\vec{k}) \ket{\Psi}_{S+G} \\
&= -\frac{ \hbar^2 }{2}   \int \frac{d^3k}{(2 \pi)^3}k^2    \hat{\tilde{h}}^T_{ij}(\vec{k})  \hat{T}^{ij}_P(-\vec{k}) \ket{\Psi}_{S+G},
\end{split}
\end{equation}
in which $\tilde{h}^T_{ij}$ is the transverse traceless part of the metric perturbation. The other double commutator is
\begin{equation}
\begin{split}
 [ \hat{\mathcal{H}}_I ,[ \hat{\mathcal{H}}_G , \hat{\mathcal{H}}_I ]] \ket{\Psi}_{S+G} & = \left [  -\frac{1}{2}\int \frac{d^3k'}{(2 \pi)^3} \hat{h}^{ij} (\vec{k}')\hat{T}^P_{ij}(-\vec{k}') , i \hbar \kappa \int \frac{d^3k}{(2 \pi)^3}\left( \hat{\pi}_{ij}^T(\vec{k})  - \frac{1}{2}P_{ij} \hat{\pi}^T(\vec{k}) \right)   \hat{T}^{ij}_P (-\vec{k})\right] \ket{\Psi}_{S+G} \\
 & =  \frac{  \kappa  \hbar^2}{2}  \int \frac{d^3k}{(2 \pi)^3} \left[\hat{T}_P^{ij} (\vec{k}) P^k_i P^l_j \hat{T}_{kl} (-\vec{k}) - \frac{1}{2}  (P_{ij} \hat{T}_P^{ij}(\vec{k}) )^2 \right]\ket{\Psi}_{S+G} \\ 
 & =  \frac{  \kappa  \hbar^2}{2}  \int \frac{d^3k}{(2 \pi)^3} \hat{\tilde{T}}^{PT}_{ij}  \hat{\tilde{T}}_{PT}^{ij} \ket{\Psi}_{S+G}.
 \end{split}
\end{equation}
In the last line, we have used a compact notation $\tilde{T}_{PT}^{ij}$ to denote the transverse traceless part of the stress-energy tensor of the  probe $P$ (we drop the hat for convenience of notation).  Note that this double commutator does not depend on specific details of the source and its corresponding quantum gravitational field. 

To obtain the final expression, it is convenient to expand the final quantum state of the full system in terms of the source energy eigenstates, since they diagonalise the Gauss constraint
\begin{equation}\label{finalstate}
\begin{split}
|\Psi_t\rangle_{SGP} &= \int d\mu(E_S)   \psi_S(E) \hat{U}_{SGP} \ket{E}_S \ket{g_E}_G \ket{\psi}_P \\
&= \int d\mu(E_S)  d\tilde{\mu}(E'_{S+G})  \psi_S(E)  \ket{E'}_S \ket{g_{E'}}_G \langle g_{E'} | \langle E' | \hat{U}_{SGP} \ket{E}_S \ket{g_E}_G \ket{\psi}_P  \\
&= \int d\mu(E_S)  d\tilde{\mu}(E'_{S+G})  \psi_S(E) \langle \hat{U}_{SGP} \rangle_{EE'}  \ket{E'}_S \ket{g_{E'}}_G \ket{\psi}_P ,
\end{split}
\end{equation}
in which we have inserted the resolution of identity in the second line. Therefore, to obtain the entangling phase from the evolution, we need to evaluate the matrix element of the time evolution operator $\langle \hat{U}_{SGP}  \rangle_{EE'}$. A requirement to observe the phase is that the experimental time scale is much smaller compared to the coherence time. Here, we truncate the expansion to the $t^3$ order and neglect higher orders in time. Note that considering higher order terms would add additional terms to the phase, but would not change the final dependence of the relative phase on the commutator of the gravitational field operators. We use a book-keeping parameter $\alpha$ (in red) to keep track of  the order of the commutator, i.e.\,we take $[\hat{h}_{ij} (\vec{k}) ,\hat{\pi}^{kl}(\vec{p}\,')]  =  i \alpha \hbar  \delta^{k}_{(i} \delta^{l}_{j)}  \delta^3(\vec{k} + \vec{p}\,')$. For the final result, we then set $\alpha =1$. Putting everything together,  we have 
\begin{equation}\begin{split}
& \langle \hat{U}_{SGP}  \rangle_{EE'} : = \langle g_E |\langle E |\hat{U}_{SGP}(t)|E'\rangle_S |g_{E'}\rangle_G  \\
 =& \langle e^{-\frac{it}{\hbar}( \hat{\mathcal{H}}_S +\hat{\mathcal{H}}_P)}  e^{- \frac{it}{\hbar}\hat{\mathcal{H}}_G }  e^{- \frac{it}{\hbar} \hat{ \mathcal{H}}_I } e^{\frac{t^2\textcolor{red}{\alpha}}{2 \hbar^2}[ \hat{\mathcal{H}}_G , \hat{\mathcal{H}}_I ]}
e^{
 \frac{i t^3 \textcolor{red}{\alpha}^2}{6 \hbar^3} \left([ \hat{\mathcal{H}}_G,[ \hat{\mathcal{H}}_G , \hat{\mathcal{H}}_I ] ]+  2[ \hat{\mathcal{H}}_I ,[ \hat{\mathcal{H}}_G , \hat{\mathcal{H}}_I ]]  \right) }  \rangle_{EE'}\\
   =& e^{-\frac{it}{\hbar} (\hat{\mathcal{H}}_P+ E_S +   E_{G})} e^
 {\frac{i t^3 \textcolor{red}{\alpha}^2 \kappa}{6 \hbar}  \int \frac{dk^3}{(2 \pi)^3} \hat{\tilde{T}}^{ij}_{PT} \hat{\tilde{T}}_{ij}^{PT}} \langle e^{ \frac{it}{2 \hbar}  \int  \frac{d^3k}{(2 \pi)^3} \hat{h}_{ij}  \hat{T}_{P}^{ij} } e^{\frac{ i \kappa  t^2 \textcolor{red}{\alpha}}{2 \hbar}   \int \frac{d^3k}{(2 \pi)^3}\left( \hat{\pi}_{ij}^T - \frac{1}{2}P_{ij} \hat{\pi}^T \right)   \hat{T}^{ij}_P }
e^{-
 \frac{i  t^3 \textcolor{red}{\alpha}^2}{12 \hbar} \int \frac{d^3k}{(2 \pi)^3}k^2  \tilde{h}^T_{ij}  \hat{T}^{ij}_P   }\rangle_{EE'} .
 \end{split}
\end{equation}
Up to this point, we have isolated the  operators of gravitational field in $\langle ....\rangle_{E E'}$ . For convenience in the calculation, we  evaluate them in the $\pi_{ij}$ basis and express the metric operator  as a functional derivative
$\hat{h}_{ij}(\vec x)  = i \hbar \frac{\delta}{\delta \pi_{ij} (\vec x)}$. Therefore, the part of the matrix element which involves gravitational operators is\footnote{One of the $\alpha^2$ order of the gravitational operator $[ \hat{H}_G,[ \hat{H}_G , \hat{H}_I] ]$ exponentiated as $\exp [-\frac{i  t^3 \alpha^2}{12 \hbar} \int \frac{d^3k}{(2 \pi)^3}k^2      \left(  \hat{h}^T_{ij}  -  \frac{P_{ij}}{2} \hat{h}^T\right)  \hat{T}^{ij}_P ]$ is omitted in this expression, since at the end it only contributes to terms at the $t^4$ order and higher after the Gaussian integration.}
\begin{equation}
\begin{split}
&\langle e^{ \frac{it}{2 \hbar}  \int  \frac{d^3k}{(2 \pi)^3} \hat{h}_{ij}  \hat{T}_{P}^{ij} }    e^{\frac{ i \kappa  t^2 \textcolor{red}{\alpha}}{2 \hbar}   \int \frac{d^3k}{(2 \pi)^3}\left( \hat{\pi}_{ij}^T - \frac{1}{2}P_{ij} \hat{\pi}^T \right)   \hat{T}^{ij}_P }\rangle_{EE'}\\
 = &  \eta^2 e^{   -\frac{i t}{4 \hbar} \int \frac{dk^3}{(2 \pi)^3} \textnormal{h}_{E'}^T  P_{ij} \hat{T}^{ij}_P }\int \mathcal{D}[\pi_{ij}]     e^{    \frac{ t \kappa}{\hbar} \int \frac{dk^3}{(2 \pi)^3} \frac{1}{|k|} \tilde{\pi}^T_{ij} \hat{T}^{ij}_P  - \frac{t^2 \kappa}{4\hbar} \int \frac{d^3 k}{(2 \pi)^3} \frac{1}{|k|} \hat{\tilde{T}}^{ij}_{PT}\hat{\tilde{T}}_{ij}^{PT}   +\frac{ i \kappa  t^2 \textcolor{red}{\alpha}}{2 \hbar}   \int \frac{d^3k}{(2 \pi)^3}\tilde{\pi}_{ij}^T   \hat{T}^{ij}_P }\Psi_{G}[\pi_{ij}, E] \Psi^*_{G}[\pi_{ij}, E'] \delta(E-E') \\
= &  e^{  - \frac{i t}{4 \hbar} \int \frac{dk^3}{(2 \pi)^3} \textnormal{h}_{E'}^T (\vec k) P_{ij} \hat{T}^{ij}_P (-\vec k)} e^{- \frac{t^2 \kappa}{4\hbar} \int\frac{d^3 k}{(2 \pi)^3}  \frac{1}{|k|} \hat{\tilde{T}}^{ij}_{PT}(\vec k) \hat{\tilde{T}}_{ij}^{PT} (-\vec k) }  e^{ \frac{ t^2 \kappa}{8 \hbar} \int \frac{d^3 k}{(2 \pi)^3} \frac{1}{|k|} \hat{T}^P_{ij}(\vec k)\hat{T}_{P}^{ij}(-\vec k)\left(1+ \frac{i t \textcolor{red}{\alpha}}{2} |k| \right)^2},
 \end{split}
\end{equation}
where we have used the same normalization condition as in \cite{Chen:2022wro} which determines $\eta$  
\begin{equation}
\eta^2 \prod_k \int d \pi^L_{ij}(\vec{k}) d\tilde{\pi}^T_{ij}(\vec{k}) d\pi^T(\vec{k})  e^ { - \frac{2 \kappa} {(2 \pi)^3 \hbar |\vec{k}| }  ( \tilde{ \pi}^T_{ij}(\vec k) + b T^P_{ij} ) ( \tilde{\pi}_T^{ij}(-\vec k)+ b T_P^{ij})  }=  \eta^2 \prod_k 2 \pi^2 \sqrt{\frac{ \hbar |k| }{ \kappa}} \int d \pi^L_{ij}(\vec{k})  d\pi^T(\vec{k})  \equiv 1.
\end{equation}
In the  Gaussian integral above, $b$  is an arbitrary function that doesn't depend on $\pi_{ij}$. In our case, $b= -\frac{t}{4}(1+ \frac{i  t \alpha}{2} |k| )$. 
Therefore, we can write down the final state Eq.\,(\ref{finalstate}) explicitly as:
\begin{equation}
\begin{split}
|\Psi_t\rangle_{SGP} 
= \int d\mu(E_S)  \psi_S(E)  & e^{-\frac{it}{\hbar} (\hat{\mathcal{H}}_P+ \hat{\mathcal{H}}_S +  \hat{\mathcal{H}}_{G})} e^{-   \frac{i t}{4 \hbar} \int \frac{dk^3}{(2 \pi)^3} \textnormal{h}_{E}^T P_{ij}\hat{T}^{ij}_P} e^
 {\frac{i t^3  \kappa \textcolor{red}{\alpha}^2}{6 \hbar}  \int \frac{dk^3}{(2 \pi)^3} \left(\hat{T}^{ij}_P   P^k_i P^l_j \hat{T}_{kl}^P  - \frac{1}{2}  (P_{ij} \hat{T}^{ij}_P)^2 \right)} \cdot \\ 
&\cdot e^{- \frac{t^2 \kappa}{4\hbar} \int\frac{d^3 k}{(2 \pi)^3}  \frac{1}{|k|} \hat{\tilde{T}}^{ij}_{PT}(\vec k) \hat{\tilde{T}}_{ij}^{PT} (-\vec k) }  e^{ \frac{ t^2 \kappa}{8 \hbar} \int \frac{d^3 k}{(2 \pi)^3} \frac{1}{|k|} T^P_{ij}(\vec k)T_{P}^{ij}(-\vec k)(1+ \frac{i t \textcolor{red}{\alpha}}{2} |k|)^2 } \ket{E}_S \ket{g_E}_G \ket{\psi}_P .
\end{split}
\end{equation}
We isolate the phase operator which depends on the interaction with the probe, which contains the relevant signature that we wish to describe. We denote as $\theta_{Free}$ the phase obtained from the free Hamiltonian of the probe, source and gravity. We additionally identify the interaction part of the phase with the operators $\hat{\Theta}^{(n)}$, with $n=0,\,1,\,2$, where the label $n$ denotes the $n-$th commutator of the gravitational field operators:
\begin{equation}
|\Psi_t\rangle_{SGP} = \int d\mu(E)\psi_S(E)  e^{-\frac{it}{\hbar}\theta_{Free}} e^{\frac{i }{\hbar}( \hat{\Theta}^{(0)} + \hat{\Theta}^{(1)} +  \hat{\Theta}^{(2)} \cdots)}\ket{E}_S \ket{g_E}_G \ket{\psi}_P ,
\end{equation}
The zeroth order $\Theta^{(0)}$ reads:
\begin{equation}\label{phizero}
    \Theta^{(0)} = - \frac{ t}{4 } \int \frac{dk^3}{(2 \pi)^3} \textnormal{h}_{E}^T (\vec k)P_{ij} \hat{T}^{ij}_P (-\vec k)-\frac{i \kappa t^2}{8 } \int \frac{d^3 k}{(2 \pi)^3} \frac{1}{|k|} \left( \hat{T}^P_{ij}(\vec k)\hat{T}_{P}^{ij}(-\vec k) - 2 \hat{\tilde{T}}^{ij}_{PT}(\vec k) \hat{\tilde{T}}_{ij}^{PT} (-\vec k) \right) .
\end{equation}
The first term is the additional entangling phase to Eq.\eqref{eq:phasewidesource}  coming from the coupling between the gravitational field and the momentum of the probe.
The second term in Eq.(\ref{phizero}) is not a phase, but an overall real damping factor.

The first order correction $\Theta^{(1)}$ to the phase coming from a single gravitational commutator is
\begin{equation} 
    \Theta^{(1)}  =  \frac{   \kappa t^3}{8 } \int \frac{d^3 k}{(2 \pi)^3}  \hat{T}^P_{ij}(\vec k)\hat{T}_{P}^{ij}(-\vec k).
\end{equation}
This quantum correction is proportional to $t^3$.  This term depends on the stress-energy tensor of the probe expressed in the frame corresponding to the temporal gauge. 

The correction $\Theta^{(2)}$ coming from  the second order of the gravitational commutator is
\begin{equation} \begin{split}
    \Theta^{(2)} & =  \frac{   \kappa t^3 }{6 }  \int \frac{dk^3}{(2 \pi)^3} \left(\hat{T}^{ij}_P   P^k_i P^l_j \hat{T}_{kl}^P  - \frac{1}{2}  (P_{ij} \hat{T}^{ij}_P)^2 \right) \\
    &= \frac{  \kappa  t^3}{6 } \int \frac{d^3k}{(2 \pi)^3} \left( \hat{\tilde{T}}^{PT}_{ij} (\vec k)  \hat{\tilde{T}}_{PT}^{ij}(-\vec k) 
    \right).
\end{split}
\end{equation}
in which the tilde $\tilde{T}_{PT}^{ij}$ denotes the transverse traceless component of $T_{P}^{ij}$. This term is also proportional to $t^3$.
Since we have kept only terms up to order $t^3$, we have neglected  the terms in $\Theta^{(1)}$ and $\Theta^{(2)}$ of the order $t^4$ and above. 

\bibliography{biblioMass}
\end{document}